\documentclass{aims}

\usepackage{txfonts}
\def\typeofarticle{Research Article}
\def\currentvolume{19}
\def\currentissue{x}
\def\currentyear{2021}

\def\ppages{xxx--xxx}
\def\DOI{}
\def\Received{Dec 2021}
\def\Accepted{XXX 2022}
\def\Published{XXX 2022}

\numberwithin{equation}{section}

\usepackage{lineno,hyperref}
\modulolinenumbers[5]
\usepackage{numcompress}
\usepackage{booktabs}
\usepackage{nomencl}
\begin{document}
\title{Fractional calculus modeling of cell viscoelasticity quantifies drug response and maturation more robustly than integer order models}

\author{%
  Anh Vo\affil{1} and Andrew Ekpenyong\affil{1,}\corrauth}
\shortauthors{the Author(s)}
\address{%
  \addr{\affilnum{1}}{Department of Physics, Creighton University, 2500 California Plaza, Omaha, NE 68178, USA}}
\corraddr{andrewekpenyong@creighton.edu; Tel: +1-402-280-2208; Fax:\\ +1-402-280-2140.}

\begin{abstract}
It has recently been discovered that the viscoelastic properties of cells are inherent markers reflecting the complex biological states, functions and malfunctions of the cells. Although the extraction of model parameters from the viscoelasticity data of many cell types has been done successfully using integer order mechanical and power-law viscoelastic models,  there are some cell types and conditions where the goodness of fits falls behind. Thus, fractional order viscoelastic models have been proposed as more general and better suited for such modeling. In this work, we test such proposed generality using published data already fitted by integer order models. We find that cell viscoelasticity data can be fitted using fractional order viscoelastic models in more situations than integer order. For macrophages, which are among the white blood cells that function in the immune system, the fractional order Kelvin-Voigt model best captures pharmacological interventions and maturation of the cells. The steady state viscosity of macrophages decreases following depolymerization of F-actin using the drug cytochalasin D, and also decreases following myosin II breakdown using Blebbistatin. When macrophages are treated with a bacterium-derived chemoattractant, the steady state viscosity decreases. Interestingly, both the steady state viscosity and elastic modulus are progressively altered as the cells become mature and approach senescence. Taken together, these results show that fractional viscoelastic modeling, more robustly than integer order modeling, enables the further quantification of cell function and malfunction, with potential diagnostic and therapeutic applications especially in cases of cancer and immune system dysfunctions. 
\end{abstract}

\keywords{
\textbf{cell mechanics, viscoelasticity, fractional calculus, fractional derivative, macrophages, cell compliance, cell deformability.}
}

\maketitle

\section{Introduction}
\label{Intro}
The mechanical properties of cells such as viscosity and elasticity have emerged as critical parameters for understanding the cell\rq{}s complex biological states, functions and malfunctions \cite{Trepat2007,Ekpenyong2012e,Chan2015}. Viscosity is a measure of a fluid\rq{}s resistance to flow when subjected to stress \cite{Mason2006}. Elasticity is a measure of a body\rq{}s ability to return to original dimensions following the application and removal of stress \cite{Lakes2009}. A viscoelastic material is therefore one that resists flow while also capable of partially returning to original dimensions after the removal of applied stress (for elaboration, see \cite{Lakes2009}). By resisting flow, viscoelastic materials dissipate mechanical energy. On the other hand, they partially return to original dimensions using stored mechanical energy. Thus, viscoelasticity is the exhibition of both viscous and elastic  properties through simultaneous dissipation and storage of mechanical energy \cite{Mainardi2010}. Biological cells are viscoelastic materials with a broad range of peculiar and adaptable features \cite{Dealy1995, Kasza2007,Trepat2007}.  
Measurements of mechanical properties of cells (and cellular constituents) such as their elasticity have been reported 
for decades \cite{Janmey1991a,Ingber1997,Stamenovic2008}. But intensive research into cell mechanics, especially as an inherent marker of functional changes is a relatively recent and burgeoning enterprise. In response to mechanical loads or stress, biological cells usually deform in different ways. Cellular mechanics or cell rheology aims at quantifying these behaviors. Such quantitative understanding can help in clarifying biological phenomena, discovering new physics (of living matter) and providing clues for diagnostic and therapeutic applications in medicine. In this work, we use both integer order and non-integer order mechanical models to extract viscoelastic parameters from strain measurements carried out on special white blood cells. Our previous work using only integer order models \cite{Ekpenyong2012e} has been well received and confirmed by several other methods of measurements and analysis within broader cell physiology contexts  \cite{Otto2015a,Man2014,Wu2018,Mierke2020}. Hence, recent developments that show the utility of fractional viscoelastic models in systematically capturing material parameters for comparison across studies \cite{Bonfanti2020,Bonfanti2020a} have directly given impetus to our extended analysis to include fractional viscoelastic models. It is pertinent to first describe stress, strain and the viscoelastic regimes we measured in order to set the stage for the modeling and characterization.

Considering stress and strain in general for a non-linear viscoelastic material, the stress, $\sigma$\nomenclature{$\sigma$}{Stress},  can be expressed as $\sigma=F[\varepsilon(t),t]$ where $\varepsilon$ \nomenclature{$\varepsilon(t)$}{Time dependent strain} is the time-dependent strain and $F$ is a functional indicating both the properties of the material and the convolving conditions of measurement such as temperature and molecular perturbations. In the linear regime, the stress and time response can be separated so that we obtain a general expression for a linear viscoelastic material as $\sigma=\varepsilon G(t)$, where $\varepsilon$ is the linear extensional strain and $G(t)$ is the time-dependent modulus of the material\nomenclature{$G(t)$}{Time dependent relaxation modulus}. Thus, in linear viscoelastic materials, the stress is proportional to the strain at all time points \cite{Findley1976}. Many non-linear viscoelastic materials have a linear regime. In this regime, the mechanical properties of the material can be separated into an elastic recoverable part  as Hooke\rq{}s law (Eq. \ref{GenHookesLaw})  and the viscous part as Newton\rq{}s law (Eq. \ref{NewtonsLaw}) \cite{Lakes2009},
\begin{subequations}
  \begin{equation}
    \sigma=E\varepsilon
    \label{GenHookesLaw}
  \end{equation}
  \begin{equation}
  \sigma= \eta\frac{d\varepsilon}{dt}.   
   \label{NewtonsLaw}
  \end{equation}
\end{subequations}
where $E$\nomenclature{$E$}{Young\rq{}s modulus} is Young\rq{}s modulus (like spring stiffness)  and $\eta$\nomenclature{$\eta$}{Coefficient of viscosity}  is the coefficient of  viscosity (like dashpot viscosity).  Obviously, such a separation is difficult in non-linear viscoelasticity. For the avoidance of confusion, we note that Eq. \ref{GenHookesLaw}  applies to linear elasticity in one dimension. The 3D version is $\sigma_{ij} = C_{ijkl}\varepsilon_{kl}$  \cite{Lakes2009}. An elastic but non-linear constitutive equation is given by 
\begin{equation}\label{nonlin}
\sigma = f(\varepsilon)\varepsilon,
\end{equation}
where $f(\varepsilon)$  is a non-linear function of strain. Both linear and non-linear elasticity have no time 
dependence. Thus, elastic materials recover fully and instantaneously when loaded or unloaded. 
Moreover, the path for loading is identical to the path of unloading, whether linear or non-linear.

For viscoelastic materials, a property that characterizes such materials when stress and strain functions are 
known is called the creep compliance, $J(t)$. In general, it is both time-dependent and stress-dependent and can therefore be denoted as $J(t,\sigma)$\nomenclature{ $J(t,\sigma)$}{Non-linear creep compliance}. Creep compliance with unit $\text{Pa}^{-1}$  is usually measured through a creep-recovery test, where the stress is kept constant and the strain is recorded as a function of time (or conversely, through a strain-relaxation test, where the strain is kept constant and stress is recorded as a function of time).  Mathematically, the constant stress in a creep recovery test can be expressed using the Heaviside step function $H(t)$, thus,
\begin{equation}
    \sigma(t)=\sigma_{0}(H(t)-H(t-t_{0})).
    \label{Heaviside}
  \end{equation}
A material initially at rest at  $t=t_{0}$, can be subjected to a constant stress $\sigma_{0}$ and the strain $\varepsilon$ monitored as a function of time. Hence, the creep compliance can be defined as $J(t,\sigma_{0})=\varepsilon(t)/\sigma$ so that using Eq. \ref{Heaviside}, we have, 
\begin{equation}
    J(t,\sigma)=\frac{\varepsilon(t)}{\sigma_{0}(H(t)-H(t-t_{0}))}
    \label{NonlinCreepComp}
  \end{equation}
In the linear regime, which can be determined experimentally \cite{Lakes2004,Lakes2009}, 
the stresses are small enough and the linear creep compliance  $J(t)$ is obtained:
\begin{equation}
  J(t,\sigma) \approx J(t).
    \label{LinCreepComp1}
  \end{equation}
Thus, using measured strain output and the known constant stress input, creep compliance for linear 
viscoelasticity is given by 
\begin{equation}
  J(t)= \frac{\varepsilon(t)}{\sigma_{0}}.
    \label{LinCreepComp2}
  \end{equation}
Creep compliance data for white blood cells are modeled in this work. The data were obtained by strain or deformation
measurements using an optical stretcher, described in detail elsewhere \cite{Guck2001,Guck2005,Wottawah2005,Ekpenyong2012e,Man2014,Wu2018} and summarized in section \ref{Experiments} on materials and methods. We have used both ``integer order'' and fractional order derivatives to model our creep compliance data for cells. In modeling creep compliance data for tissues and tissue-like materials, it was shown that the fractional calculus (FC) models were better in fitting data \cite{Meral2010}, provoking the need to attempt this for cells. A recent review of the role of FC in modeling biological phenomena \cite{Ionescu2017} shows increasing use of FC for more accurate modeling of diffusion of substances in the human body at the system and organ levels, and for viscoelasticity of tissues and action potentials in neurons. Attempts at standardization and generalization of viscoelastic characterization using FC, aimed at widespread adoption have emerged \cite{Bonfanti2020,Bonfanti2020a}. Our work here shows that FC models accurately quantify the response of specific immune cells (here macrophages), to cytoskeletal drugs and chemical agents that signal bacterial infections, as well as their cellular maturation, just like the integer order models. Such quantification have implications for disease diagnosis and therapeutic options especially in the case of cancer, where alterations in cell mechanics not only indicate malignant transformation \cite{Guck2001,Guck2005,Lautenschlager2009,Gossett2012} but have been identified as therapeutic target \cite{Guck2013,Prathivadhi-Bhayankaram2016d}. 
\section{Materials and methods}
\label {Experiments}
\subsection{Measurement of viscoelastic properties of cells}
In the optical stretcher, forces generated by the momentum transfer from two non-focused,
counter-propagating, divergent laser beams to the surface of single suspended cells are employed to 
trap and deform the cells in a controlled and non-destructive way, thereby enabling the extraction of cell 
viscoelastic properties. Very instructive comparative reviews of most of the other methods of measuring cell mechanical properties  
can be found in \cite{VanVliet2003,Guck2010,Wu2018}. There, micropipette aspiration, microplate manipulation, 
optical stretching and microfluidic deformation are evaluated together as techniques that measure 
global mechanical properties of cells. Furthermore, a different review \cite{DiCarlo2012} and designed experimental comparison \cite{Wu2018} have been particularly useful in categorizing the techniques based on envisaged translation to clinical applications. 

Let us focus on creep compliance as measured by the optical stretcher, OS. During a typical OS experiment, cells are introduced into a 
microfluidic delivery system, serially trapped and then stretched along the laser beam axis. The elongation of the cell body along the 
laser beam axis is recorded by a CCD camera. The time-dependent strain is extracted from the video camera images \cite{Ekpenyong2012e}. The axial strain during optical stretching is given by
\begin{equation}\label{axial_strain}
\varepsilon =\frac{a(t)}{a_{0}}-1,
\end{equation}
where $a_{0}$ is the semi-major axis of the unstretched cell and $a(t)$ is the time-varying semi-major axis measured. The optically induced surface stress on the cells is computed using an electromagnetic wave model \cite{Boyde2009,Boyde2012a}, which requires knowledge of the average refractive index of cells. The average refractive index of cells was measured using a digital holographic microscope \cite{Chalut2012,Ekpenyong2013}. Thus, the mechanical properties of cells obtained from OS measurements are decoupled from their optical characteristics. The tensile strain is normalized by the peak value of the calculated optical stress $\sigma_{0}$ and a geometric factor $F_{g}$ to give the tensile compliance for each cell:
\begin{equation}
  J(t)= \frac{\varepsilon(t)}{\sigma_{0}F_{g}}.
    \label{LinCreepComp3}
  \end{equation}
Eq. \ref{LinCreepComp3} is identical to Eq. \ref{LinCreepComp2} except for the dimensionless $F_{g}$, which is calculated as described elsewhere \cite{Wottawah2005a} to account for cell shape and stress distribution. 
Insights into the often complex response of viscoelastic materials to stress or strain can be gained using appropriate assemblies of simple springs and dash pots which reflect the elastic and viscous aspects respectively. Also, scaling laws such as power law models can be used to analyze cellular viscoelastic responses \cite{Ekpenyong2012c,Chan2015}. Relevant mechanical models are presented in the theory and calculation section. We next describe the intrinsic bases of the viscoelastic properties of cells, namely, cytoskeletal structures and how we altered these in order to obtain viscoelastic readouts.  


\subsection{Alterations of cytoskeletal structures and cell states}
The typical eukaryotic cell  has evolved structures in the cytoplasm and in the nucleus that enable it to perform the following 
mechanically relevant functions: spatial organization of its contents; physical and biochemical connection 
to the external environment; and generation of coordinated forces that enable it to move and change 
shape \cite{Fletcher2010}. These cytoskeletal structures, collectively called cytoskeleton, and the 
nuclear structures  integrate the activity of a multitude of  proteins and organelles. Let us focus on the cytoskeleton. 
The eukaryotic cytoskeleton consists of three main polymers: actin filaments (or microfilaments), 
microtubules and intermediate filaments, all organized into networks whose architecture  
is controlled by several classes of regulatory proteins. These three groups of cytoskeletal 
polymers differ in their mechanical stiffness, the dynamics of their assembly (polymerization), 
their polarity and their associated molecular motors \cite{Fletcher2010} and so their continuous
remodeling leads to changes in the viscoelastic properties of cells. 

Having obtained macrophages by the differentiation of HL60 cells using a method described elsewhere \cite{Ekpenyong2012e,Ekpenyong2012c}, we used specific drugs and chemical agents to  perturb cytoskeletal structures thereby altering cell states and then measured the viscoelastic properties as readouts of such alterations.  The drugs include blebbistatin (Blebb), which inhibits myosin II and cytochalasin D (CytoD), which depolymerizes filamentous actin in the cell. Details of these pharmacological interventions are reported in our previous work \cite{Ekpenyong2012e,Chan2015,Ekpenyong2017b}. Furthermore, macrophages detect the presence of bacteria using chemical compounds produced by the bacteria, such as N-formylmethionine leucyl-phenylalanine (fMLP). Thus, we tested whether cellular viscoelastic properties are involved in cell function by measuring changes in compliance due to fMLP treatment \cite{Ekpenyong2012e}. All these methods involved specific alterations of the cytoskeleton and the mechanical make-up of the cell. The biomedical significance of these experiments and results engenders the need for further analysis, simulations,  curve-fitting and modeling which we now describe in detail in the next section. 
\subsection{Modeling}
\subsubsection{``Integer-Order'' Mechanical models}
Mechanical elements such as linear springs (for elastic response) and linear dashpots (for viscous response) can be combined to form 
simple models of the viscoelastic properties of materials.  Although the models do not convey information about the molecular or microscopic processes going on in the material, they can serve as phenomenological guides giving insight into the nature of the viscoelastic response of materials. Cells respond to constant creep stress in a viscoelastic manner. The input is the stress and the output is the strain. The strain profile can be more or less viscous. Models are very useful for extracting viscoelastic parameters from strain or deformation measurements. We consider six linear mechanical models of creep compliance to choose from, in fitting our experimental data. The equilibrium conditions and constitutive equations for deriving these models are straightforward and details can be found in standard rheology texts \cite{Findley1976,Lakes2009}. Here, we indicate briefly the mathematical essence of such derivations to aid the acquisition of physical intuition about the resulting equations and models.

\begin{figure}[!htbp]
  \centering
    \leavevmode
      \includegraphics[scale=1]{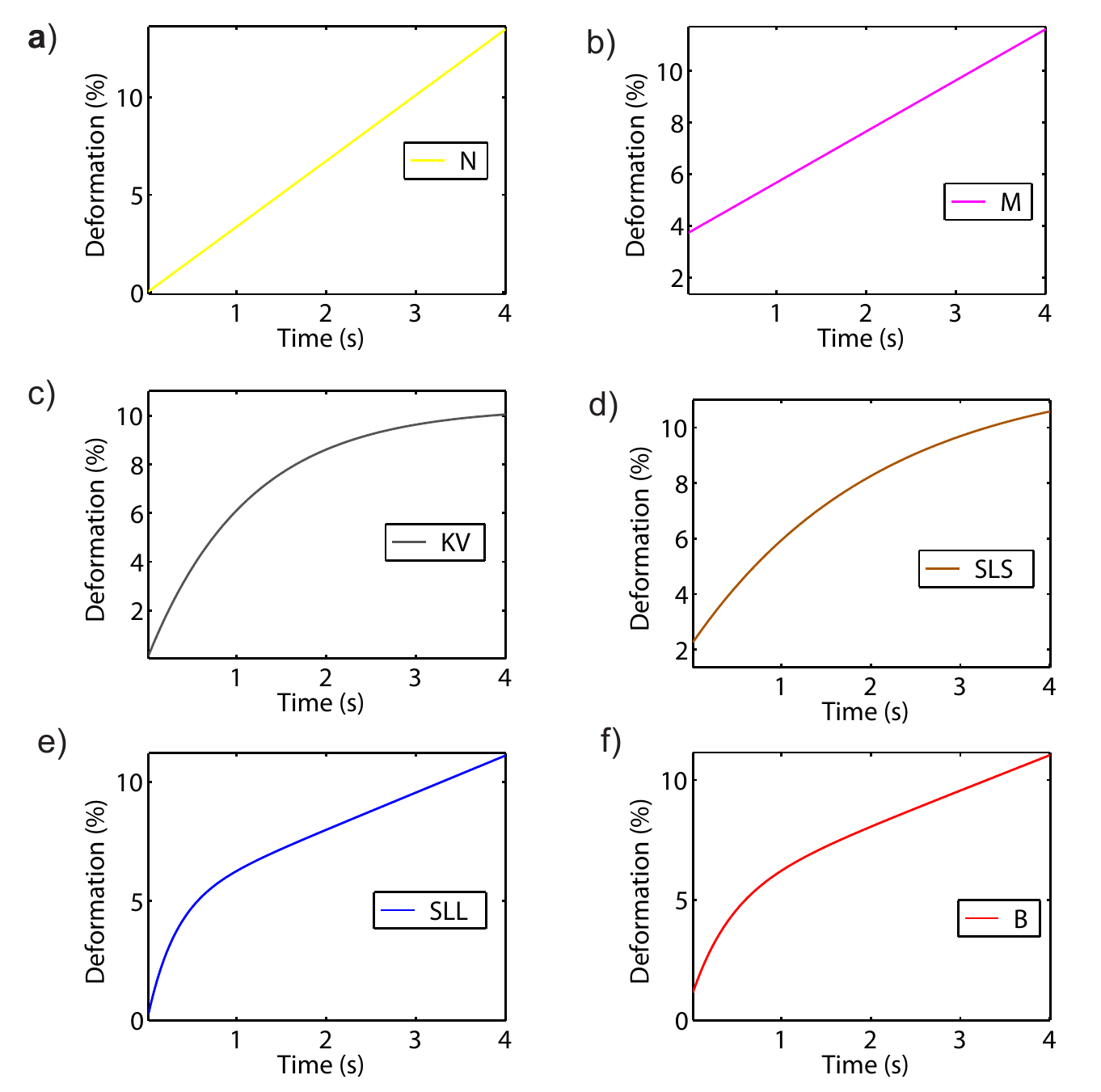}
    \caption[Common mechanical models]{The creep part of the strain or deformation for all six mechanical models. a) Newton model, N. b) Maxwell model, M. c) Kelvin-Voigt model, KV. d) Standard linear solid model, SLS. e) Standard linear liquid, SLL. f) Burgers model, B. The models were plotted using the same input parameter, namely, a strain function of time ranging from 0 to 11\% for 4 s, chosen as an example giving strain/compliance shapes typical of the cells we have measured.}
    \label{Figall_mech_models}
 \end{figure}

First, the Newton model (N),\nomenclature{N}{Newton\rq{}s model} consisting of one dashpot, $\eta_{1}$ has creep compliance given by,
\begin{equation}
  J(t)= \frac{1}{\eta_{1}}t.
    \label{N}
  \end{equation}
The N model describes the creep compliance of a perfectly viscous body. Eq. \ref{N} is obtained by replacing $\sigma(t)$ in Eq. \ref{NewtonsLaw} with $\sigma_{0}$ and integrating with respect to time:
\begin{subequations}
  \begin{equation}
    \sigma_{0}=\eta\frac{d\varepsilon}{dt}.   
    \label{N1}
  \end{equation}
  \begin{equation}
  \sigma_{0}\int\,\mathrm{d}t= \eta\int\,\mathrm{d}\varepsilon = \eta\varepsilon + c,   
   \label{N2}
  \end{equation}
\end{subequations}
where $c$ is a constant of integration to be determined by initial conditions. For the N model, $c$ is 0.
Other models are derived from the equilibrium conditions and constitutive relations by similar substitution of the $\sigma_{0}$ and direct integration. Next is the so-called Maxwell model (M), \nomenclature{M}{Maxwell\rq{}s model}made up of a spring and a dashpot in series, such that the creep compliance can be expressed as
\begin{equation}
  J(t)= \frac{1}{E_{1}}+\frac{1}{\eta_{1}}t.
    \label{M}
  \end{equation}
In the M model, the constant of integration (Eq. \ref{N2}) is not 0 because there is an instantaneous elasticity provided by the Hookean spring. Third is the Kelvin model which  is also called the Voigt model. Throughout this work, we denote it as the Kelvin-Voigt model (KV). \nomenclature{KV}{Kelvin-Voigt model}The KV model has a spring and dashpot in parallel. Its creep compliance is denoted thus:
\begin{equation}
  J(t)= \frac{1}{E_{2}}\left(1-e^{-\frac{E_{2}}{\eta_{2}}t}\right).
    \label{KV}
  \end{equation}
A spring and a KV element combined in series give the so-called Standard Linear Solid (SLS) or Zener model \nomenclature{SLS}{Standard Linear Solid or Zener model} with creep compliance given by
\begin{equation}
  J(t)= \frac{1}{E_{1}}+\frac{1}{E_{2}}\left(1-e^{-\frac{E_{2}}{\eta_{2}}t}\right).
    \label{SLS}
  \end{equation}
When a dashpot is linked in series with a KV element, we get the Standard Linear Liquid model (SLL)\nomenclature{SLL}{Standard Linear Liquid or Anti-Zener model} whose creep compliance has the form
\begin{equation}
  J(t)= \frac{1}{\eta_{1}}t+\frac{1}{E_{2}}\left(1-e^{-\frac{E_{2}}{\eta_{2}}t}\right).
    \label{SLL}
  \end{equation}
A sixth mechanical model  is called the Burgers model (B), \nomenclature{B}{Burgers\rq{} model} consisting of a KV element and a Maxwell model in series. The creep compliance is expressed as
\begin{equation}
  J(t)= \frac{1}{E_{1}}+\frac{1}{\eta_{1}}t+\frac{1}{E_{2}}\left(1-e^{-\frac{E_{2}}{\eta_{2}}t}\right).
    \label{B}
  \end{equation}
The creep part of the strain or deformation (from which the creep compliance is derived by simple normalization) for the Burgers model and all the other five mechanical models are illustrated in Fig. \ref{Figall_mech_models}. 
In view of a transition to fractional order models, we state explicitly the constitutive equations for N, M, KV, SLS, SLL and B.
For N, the constitutive equation is  

  \begin{equation}
   \sigma(t) =  b_{1}\frac{d}{dt}\varepsilon(t),
      \label{CN}
    \end{equation}
  where $b_{1}=\eta_{1}$, when this equation is solved for the creep case when $\sigma(t) = \sigma_{0}$. The constants $a_{k}$ and $b_{k}$ are determined for each model accordingly. The constitutive equation for M is

  \begin{equation}
   \sigma(t) + a_{1}\frac{d\sigma}{dt} =  b_{1}\frac{d}{dt}\varepsilon(t).
      \label{CM}
    \end{equation}

  It is 

  \begin{equation}
   \sigma(t) = m\varepsilon(t)+ b_{1}\frac{d}{dt}\varepsilon(t),
      \label{CKV}
    \end{equation}
  for the KV model.

  For SLS, it is

  \begin{equation}
  \left(1+ a_{1}\frac{d}{dt}\right)\sigma(t) =  \left(m+ b_{1}\frac{d}{dt}\right)\varepsilon(t).
      \label{CSLS}
    \end{equation}
  And for SLL, we have

  \begin{equation}
  \left(1+ a_{1}\frac{d}{dt}\right)\sigma(t) =  \left(b_{1}\frac{d}{dt} + b_{2}\frac{d^2}{dt^2}\right)\varepsilon(t).
      \label{CSLL}
    \end{equation}

The constitutive equation for the Burgers\rq{} 4-element model is

\begin{equation}
  \sigma(t)+\left(\frac{\eta_{1}}{E_{1}} + \frac{\eta_{1}}{E_{2}}+\frac{\eta_{2}}{E_{2}}\right)\dot\sigma(t) + \frac{\eta_{1}\eta_{2}}{E_{1}E_{2}}\ddot\sigma(t)=\eta_{1}\dot{\varepsilon}(t) + \frac{\eta_{1}\eta_{2}}{E_{2}}\ddot\varepsilon(t)
    \label{CB}
  \end{equation}

It is straightforward to rename the elastic and viscous constants in order to obtain a compact expression:
\begin{equation}
  \left(1+ a_{1}\frac{d}{dt} + a_{2}\frac{d^{2}}{dt^{2}}\right)\sigma(t) = \left(b_{1}\frac{d}{dt} + b_{2}\frac{d^{2}}{dt^{2}}\right)\varepsilon(t),
    \label{CBB}
  \end{equation}

where $a_{1} = \left(\frac{\eta_{1}}{E_{1}} + \frac{\eta_{1}}{E_{2}}+\frac{\eta_{2}}{E_{2}}\right)$,  $a_{2}=\frac{\eta_{1}\eta_{2}}{E_{1}E_{2}}$, $b_{1}=\eta_{1}$ and $b_{2}=\frac{\eta_{1}\eta_{2}}{E_{2}}$. \\

Clearly, Eq.\ref{CBB} has the following general form
\begin{equation}
 \left(1+\sum_{k=1}^p a_{k}\frac{d^{k}}{dt^{k}}\right)\sigma(t) =  \left(m+\sum_{k=1}^qb_{k}\frac{d^{k}}{dt^{k}}\right)\varepsilon(t), 
    \label{CBG}
  \end{equation}
  where, for the Burgers\rq{} model $m=0$ and $p=q=2$. At this stage, $p$ and $q$ are integers and to meet physical requirements, $q=p$ or $q=p+1$, while $a_{k},\;b_{k}$ and $m$ are non-negative constants. For N, $m=0$, $p=0$ and $q=1$, for M, $m=0$ and $p=q=1$, for KV, $m>0$ and $p=0\;q=1$, for SLS, $m>0$ and $p=q=1$ and for SLL, $m=0$, $p=1$ and $q=2$. Thus, we can use Eq. \ref{CBG} to generate all the material constants for the SLL, SLS, KV, M, and N models, and also have it as a synthesis of the various formalisms (integral, differential, matrices and geometric construction) used in obtaining the models. We next use  Eq. \ref{CBG} for a transition to fractional calculus-based models of viscoelasticity. Anecdotally, there are  alternatives to mechanical models, other than fractional element models. Power law models are among such alternatives which we also used in \cite{Ekpenyong2012e}.
Only the overlapping insights from both power-law and mechanical models impinge upon the conclusions of our work, hence, our work does not address the ongoing debate as to whether cellular viscoelastic behaviors truly follow power-laws \cite{Kollmannsberger2010,Lenormand2004} or mechanical models \cite{Karcher2003,Wottawah2005a}. Rather, we present our finding that fractional calculus modeling of cellular viscoelasticity is a general framework for characterizing cellular behavior, a framework of which both integral order mechanical models and integral order power law models are but parts, as has been reviewed recently \cite{Bonfanti2020}. 

\subsubsection{Fractional calculus models}
  Consider the constitutive equation for the Burgers\rq{} 4-element model (Eq.\ref{CBG}). Without undermining linearity and without any loss of generality, Eq.\ref{CBG} can be expressed as

  \begin{equation}
   \left(1+\sum_{k=1}^p a_{k}\frac{d^{\nu_{k}}}{dt^{\nu_{k}}}\right)\sigma(t) =  \left(m+\sum_{k=1}^qb_{k}\frac{d^{\nu_{k}}}{dt^{\nu_{k}}}\right)\varepsilon(t), 
      \label{CBG2}
    \end{equation}
  where $\nu \in{0,1}$ and $\nu_{k}=k+\nu -1$. As long as $\nu$ is either 0 or 1, then we have integer order derivative in Eq.\ref{CBG2}. Also, Eq.\ref{CBG2} is the constitutive relation for linear viscoelasticity in differential form. It\rq{}s integral form usually adopts the so-called Boltzmann superposition principle to present strain as the result of stress. Let us focus on the differential form of the constitutive relation (Eq.\ref{CBG2}). If $\nu$ is allowed to be continuous between 0 and 1, then we have fractional order derivatives. This plunges us into the sea of fractional calculus, engendering fractional element modeling of cellular viscoelasticity. More explicitly, within the linear viscoelastic regime, as we showed in the introduction, the mechanical properties of the material can be separated into an elastic recoverable part  as Hooke\rq{}s law (Eq. \ref{GenHookesLaw}) represented by a spring, and the viscous part as Newton\rq{}s law (Eq. \ref{NewtonsLaw}) \cite{Lakes2009}, represented by a dashpot. Fractional order derivatives allow us to combine the spring and the dashpot into the so-called spring-pot \cite{Meral2010} thus:
  \begin{subequations}
    \begin{equation}
      \sigma(t)=E\frac{d^{0}\varepsilon}{dt^{0}}\; (spring)
      \label{FCGenHookesLaw}
    \end{equation}
    \begin{equation}
    \sigma(t)= \eta\frac{d^{1}\varepsilon}{dt^{1}}\; (dashpot)
     \label{FCNewtonsLaw}
    \end{equation}
   \begin{equation}
    \sigma(t)= \eta\frac{d^{\nu}\varepsilon}{dt^{\nu}}\; (springpot),
     \label{Springpot}
    \end{equation}
  \end{subequations}
  where $\nu$ is allowed to vary continuously between 0 and 1. Almost prereflexively, we can glimpse fractional order mechanical models as series/parallel combinations of springs and springpots. For historical reasons, springpots are called Scott-Blair elements \cite{Loverro2004,Mainardi2010}.  

Using springpots (Eq. \ref{Springpot}) the fractional order constitutive equation for  B can be written as a further generalization of \ref{CBG2}:
  \begin{equation}
    (1+ a_{1}\frac{d^{\nu}}{dt^{\nu}} + a_{2}\frac{d^{2}}{dt^{1+\nu}})\sigma(t) = (b_{1}\frac{d^{\nu}}{dt^{\nu}} + b_{2}\frac{d^{1+\nu}}{dt^{1+\nu}})\varepsilon(t),
      \label{FCB2}
    \end{equation}

  Details of the derivation of the fractional order constitutive equations and compliance for the models we are using (N, M, KV, SLL, SLS and B) can be found in \cite{Mainardi2011,Dalir2010,Zhou2011}. We have followed closely the notation in  \cite{Mainardi2011}, in order not to worsen the current situation in the fractional calculus field where diverse notations and formalisms tend to obfuscate the underlying mathematical unity. The fractional order expressions of strain for B, SLL, SLS, KV and M can all be found in \cite{Mainardi2010}, where Mainardi in particular has derived them with rigor and elegance. We used these expressions in fitting the compliance data. 
\subsubsection{Implementation and curvefitting in Matlab}
Caputo and Mainardi connected linear viscoelastic models with fractional order models back in the early 1970s \cite{Caputo1971,Loverro2004} and introduced the following very useful correspondence principles which make it less laborious to move from classical viscoelastic models to fractional order models \cite{Caputo1971,Mainardi2010}:
  \begin{equation}
    0< \nu < 1 
  \left\{ \begin{array}{lll}
      t\rightarrow \frac{1}{\Gamma(1+\nu)}{\left(\frac{t}{\tau_{0}}\right)}^{\nu}&\\
       \delta(t)\rightarrow \frac{1}{\Gamma(1-\nu)}\left(\frac{t}{\tau_{0}}\right)^{-\nu}&\\
      e^{-\frac{t}{\lambda}} \rightarrow E_{\nu}\left(-{\frac{t}{\lambda}}^{\nu}\right),
         \end{array} \right.
      \label{correspondence}
    \end{equation}
  where $\lambda$ denotes the relaxation time, $E_{\nu}$ is a generalized exponential function called the Mittag-Leffler function (mlf) (see \cite{Mainardi2011} for details), here, of order $\nu$ and $\delta(t)$ is the Dirac delta distribution. Mainardi and Spada \cite{Mainardi2011} have provided transparent expressions for both the mechanical models and their FC versions ready for implementation in programs such as MATLAB. We used their expressions \cite{Mainardi2011} for fractional versions of M, KV, SLS and SLL in the case of creep for fitting cell compliance data. For instance, the FC version of KV (Frac KV) is represented by a Hookean spring and a  springpot or Scott-Blair element. The Frac KV constitutive relation is therefore
   \begin{equation}
    \sigma(t)=E\varepsilon(t)+ \eta\frac{d^{\nu}\varepsilon}{dt^{\nu}}\;, 
     \label{CFracKV}
    \end{equation}
where for clarity, $m$ and $b_{1}$ in Eq \ref{CKV} are the the elastic modulus and coefficient of viscosity, respectively. The creep compliance for the Frac KV, following the notation in \cite{Mainardi2011} and our expression for KV in Eq \ref{KV}, is 
\begin{equation}
  J(t)= \frac{1}{E_{2}}\left(1-E_{\nu}\left(-{\frac{E_{2}}{\eta_{2}}{t}^{\nu}}\right)\right).
    \label{FracKV}
  \end{equation}
Thus, we are able to make direct comparisons between the parameters from the integer order mechanical models and their FC versions. The MATLAB function for the mlf is publicly available \cite{Podlubny2012}. Using the mlf, we developed an in-house MATLAB routine that implements least-squares curve-fitting for both integer order and fractional order mechanical models. 

 \section{Results and discussion}
\subsection{Integer order mechanical models fit cell compliance data}
\begin{figure}[!htbp]
\centering
    \leavevmode
      \includegraphics[scale=0.6]{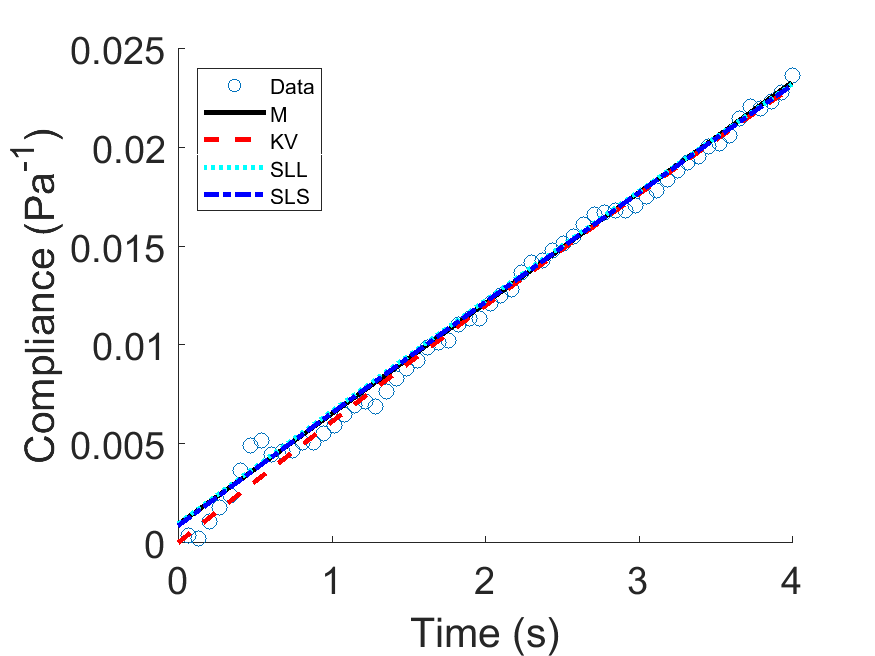}
       \caption[Integer order mechanical models]{Integer order mechanical models. The average creep compliance of macrophages are fitted using four models, M, KV, SLS and SLL.}
    \label{FigInteger_Mech_Models}
\end{figure}

Commonly used integer order mechanical models do in fact fit cell compliance data well, including the compliance data of macrophages as we show in Fig. \ref{FigInteger_Mech_Models}. 
Each of the models produced meaningful parameters (see Table \ref{TabInteger_Mech_Models}) for characterizing the viscoelastic properties of macrophages. As shown in Table \ref{TabInteger_Mech_Models}, in terms of producing physically meaningful parameters as well as the goodness of fit parameters which are relevant here for comparing the models, the SLL is the best of the four mechanical models. Note that the goodness of fit parameters are the sum of squared errors (sse), coefficient of determination (rsquare), adjusted coefficient of determination (adjusted rsquare) and the  root-mean-square error (rmse). In all the tables, dfe refers to the degrees of freedom (number of cells fitted). 
\begin{table}[H]
\centering
\caption{Parameters of integer order mechanical models}
\label{TabInteger_Mech_Models}
\begin{tabular}{@{}lllll@{}}
\toprule
Parameters      & M           & KV          & SLL         & SLS         \\ \midrule
$E_1$            & 1081.47 & 0 & 0  & 1166.70 \\
$\eta_{1}$      & 178.65  & 0  & 1038.97 & 0 \\
$E_2$            & 0           & 6.33        & 2.64           & 174.19  \\
$\eta_{2}$         & 0           & 160.38         & 174.38 & 2.50         \\
sse             & 1.7645E-05 & 1.4958E-05 & 2.0961E-05 & 1.8046E-05 \\
rsquare         & 0.9931 & 0.9942 & 0.9918 & 0.9930 \\
dfe             & 57          & 57          & 56          & 56          \\
adjusted rsquare & 0.9930 & 0.9941 & 0.9915  & 0.9927 \\
rmse            & 0.0006 & 0.0005  & 0.0006 & 0.0006 \\ \bottomrule
\end{tabular}
\end{table}
\subsection{Partial failure of integer order mechanical models}
However, when the viscoelastic properties of the macrophages were carefully altered using drugs and following their maturation (at 96 hours or h96), none of the integer models fitted all the data while producing meaningful parameters (see Table \ref{TabBad_Int_Mech_Models}). 
\begin{figure}[!htbp]
\centering
    \leavevmode
      \includegraphics[scale=0.6]{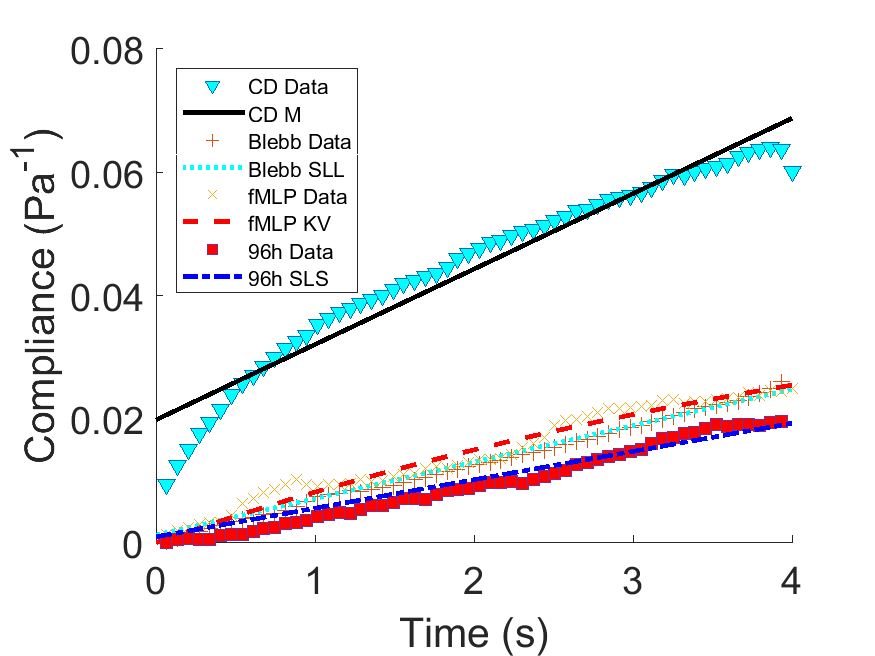}
       \caption[Failures of integer order mechanical models]{Failures of integer order mechanical models. Instances where the four models, M, KV, SLS and SLL, failed to fit and produce meaningful parameters. M failed when macrophages where treated with CD. For Blebb-treated macrophages, SLL is a poor fit. For fMLP-treated cells, KV is a poor fit. For more mature macrophages (96h), SLS is a poor fit.}
    \label{FigBad_Int_Mech_Models}
\end{figure}
Each had limitations as shown in Fig. \ref{FigBad_Int_Mech_Models}.  When macrophages were treated with cytochalasin D (CD) which depolymerizes filamentous actin in the cell, the M model clearly failed to include data from the first 0.6 seconds, rendering the fit spurious, even from a cursory visual inspection. For macrophages treated with blebbistatin (Blebb) which inhibits myosin II, SLL is a poor fit. For fMLP-treated cells, KV is a poor fit. For more mature macrophages (96h), SLS is a poor fit. These limitations are further seen in the table of parameters (Table \ref{TabBad_Int_Mech_Models}) where rmse increases, rsquare and adjusted rsquare decrease reflecting the poor fitting seen in the figure (Fig. \ref{FigBad_Int_Mech_Models}). Moreover, the results were not stable during repeated fitting routines, as were those of Table \ref{TabInteger_Mech_Models}.
\begin{table}[]
\centering
\caption{Parameters for failures of integer order mechanical models}
\label{TabBad_Int_Mech_Models}
\begin{tabular}{@{}lllll@{}}
\toprule
Parameters      & CD\_M       & fMLP\_KV    & Blebb\_SLL  & h96\_SLS    \\ \midrule
$E_1$            & 50.19 & 0 & 0 & 1027.58 \\
$\eta_{1}$          & 81.86 & 0 & 797.36 & 0 \\
$E_2$            & 0           & 20.21       & 0.12           & 215.15 \\
$\eta_{2}$         & 0           & 111.30           & 168.95 & 1.08          \\
sse             & 0.0007 & 0.0001 & 2.5368E-05 & 0.0001 \\
rsquare         & 0.9473 & 0.9650 & 0.9914 & 0.9516  \\
dfe             & 57          & 57          & 55          & 55          \\
adjusted rsquare & 0.9464 & 0.9643 & 0.9911 & 0.9498\\
rmse            & 0.0034 & 0.0014  & 0.0007 & 0.0014  \\ \bottomrule
\end{tabular}
\end{table}
\subsection{FC mechanical models fit cell compliance data}
\begin{figure}[!htbp]
\centering
    \leavevmode
      \includegraphics[scale=0.6]{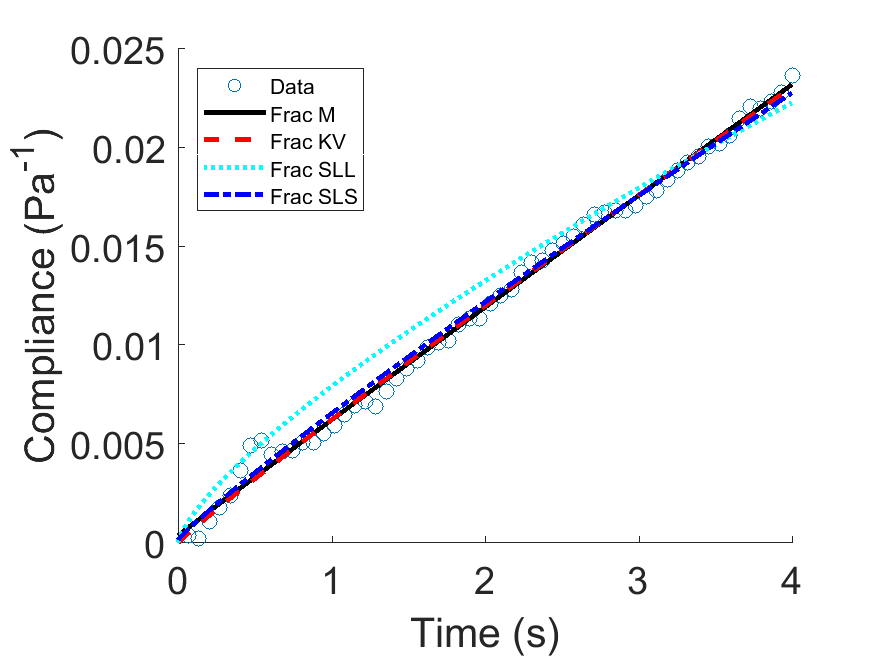}
       \caption[FC mechanical models]{Fractional calculus mechanical models. The average creep compliance of macrophages are fitted using four FC models, Frac M, Frac KV, Frac SLS and Frac SLL.}
    \label{FigFC_Mech_Models}
\end{figure}

Using the same macrophage compliance data as in  Fig. \ref{FigInteger_Mech_Models}, we found that the FC mechanical models fit the data well, as shown in Fig. \ref{FigFC_Mech_Models}. Based on the table of viscoelastic parameters and the goodness of fits (Table \ref{TabFC_Mech_Models}), the Frac KV model is the best for this set of macrophage data. It has the best adjusted rsquare, along with M, but unlike M, the viscoelastic parameters are stable of several iterations. The instability in using M is shown in the elastic modulus from M, with five orders of magnitude (40977 Pa). 
 \begin{table}[]
\centering
\caption{Parameters from fractional calculus mechanical models }
\label{TabFC_Mech_Models}
\begin{tabular}{@{}lllll@{}}
\toprule
Parameters      & Frac\_M     & Frac\_KV    & Frac\_SLL   & Frac\_SLS   \\ \midrule
$E_1$            & 40977.27 & 0  & 0 & 3168.98 \\
$\eta_{1}$         & 164.23 & 0 & 127.95 & 0 \\
$E_2$            & 0           & 0.91           & 8184.51           & 164.67 \\
$\eta_{2}$       & 0           & 163.23          & 591.95 & 2.28          \\
$\nu$               & 0.94  & 0.95 & 0.66 & 0.96  \\
sse             & 1.4117E-05 & 1.4222E-05 & 0.0002 & 1.5576E-05 \\
rsquare         & 0.9945 & 0.9944 & 0.9249 & 0.9939 \\
dfe             & 56          & 56          & 55          & 55          \\
adjusted rsquare & 0.9942  & 0.9942 & 0.9208 & 0.9936 \\
rmse            & 0.0005 & 0.0005 & 0.0019 & 0.0005 \\ \bottomrule
\end{tabular}
\end{table}

\subsection{FC mechanical models quantify cytoskeletal changes}
\begin{figure}[!htbp]
\centering
    \leavevmode
      \includegraphics[scale=0.6]{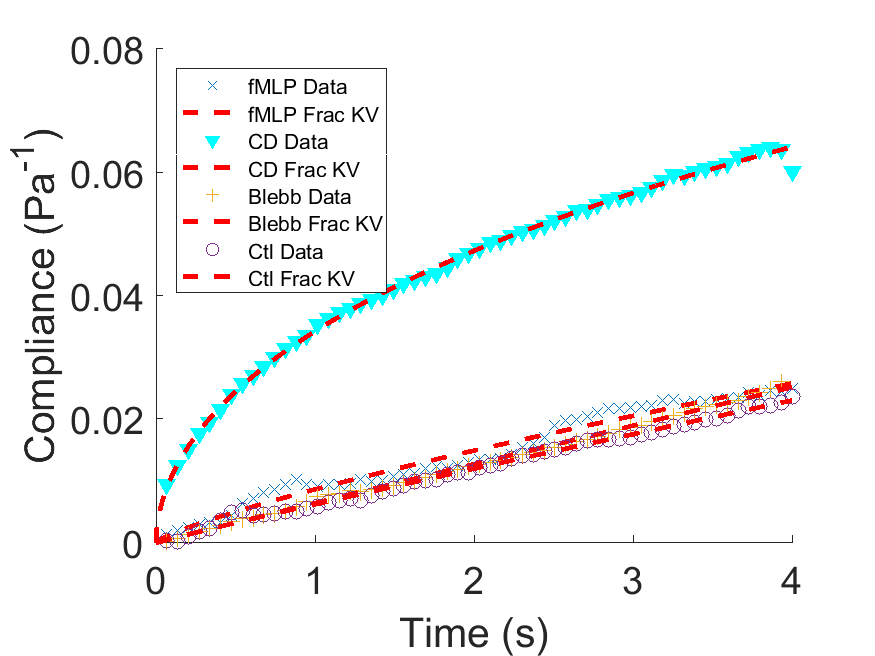}
       \caption[FC KV model quantifies drug-induced changes]{Fractional calculus KV model quantifies drug-induced changes in cell viscoelastic properties. The drugs CD, Blebb and the chemoattractant fMLP have known specific effects on cell cytoskeleton.}
    \label{FigFC_KV_Drugs}
\end{figure}

Having found that Frac KV is the best in fitting compliance data of macrophages, we went further to use this model to fit compliance data for macrophages subjected to pharmacological interventions. Interestingly, we also found that Frac KV  fits well and produces meaningful parameters that quantify drug-induced alterations of cell viscoelastic properties as shown in Fig. \ref{FigFC_KV_Drugs}.
The parameters enable viscoelastic characterization of drug responses. In Table \ref{TabFC_KV_Drugs}, the elastic and viscous constants change in drug-dependent fashion since these drugs have generic effects on cells depending on whether they are adherent or in suspension. Our macrophages were measured in suspension \cite{Ekpenyong2012e,Chan2015}. The steady state viscosity, $\eta_{2}$, decreases from 163 Pa.s to 29 Pa.s following the treatment of macrophages with cytochalasin D, which means a reduction in the cell's resistance to flow. Not surprisingly, the elastic modudus, $E_2$, increases by an order of magnitude from 0.91 Pa to 3.88 Pa. With blebbistatin treatment, Frac KV model reveals a decrease in elastic modulus and a decrease in steady state viscosity. These results are in consonance with the already noted peculiar responses of macrophages to blebbistatin, compared to many other adherent but suspended cells \cite{Chan2015}.
\begin{table}[]
\centering
\caption{Frac KV model parameters that quantify drug-induced changes}
\label{TabFC_KV_Drugs}
\begin{tabular}{@{}lllll@{}}
\toprule
Parameters      & Ctl\_Frac\_KV & Blebb\_Frac\_KV & CD\_Frac\_KV & fMLP\_Frac\_KV \\ \midrule
$E_2$               & 0.91    & 0.17     & 3.88  & 0.32    \\
$\eta_{2}$            & 163.23   & 155.52     & 29.47  & 125.13    \\
$\nu$               & 0.95   & 0.98     & 0.53  & 0.80    \\
sse             & 1.4222E-05   & 8.9647E-06     & 2.8295E-05  & 9.0973E-05    \\
rsquare         & 0.9944   & 0.9970     & 0.9977  & 0.9695    \\
dfe             & 56            & 55              & 56           & 56             \\
adjusted rsquare & 0.9942   & 0.9968     & 0.9976  & 0.9684    \\
rmse            & 0.0005   & 0.0004     & 0.0007  & 0.0013    \\ \bottomrule
\end{tabular}
\end{table}
\begin{figure}[!htbp]
\centering
    \leavevmode
      \includegraphics[scale=0.6]{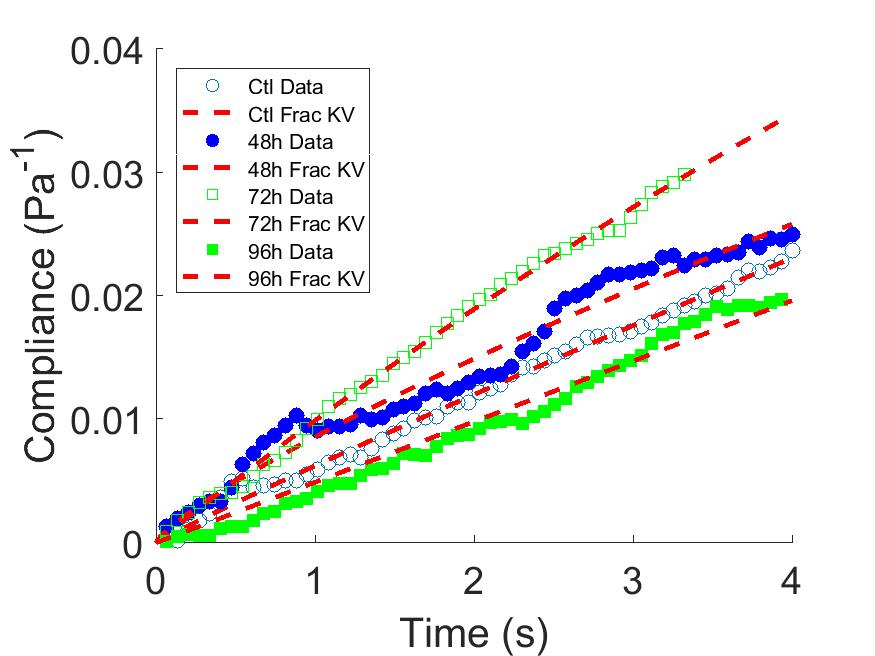}
       \caption[Frac KV model quantifies cell cycle changes]{Fractional calculus KV model quantifies changes in cell viscoelastic properties due to cell cycling and maturation. From 24 h following induction of differentiation to 96 h, the macrophages exhibit cell cycling induced changes in their cytoskeleton apparent in the fit parameters}
    \label{FigFC_Time}
\end{figure}
\begin{table}[!htbp]
\centering
\caption{Parameters from Frac KV model reflecting cell cycling and maturation}
\label{TabFC_Time}
\begin{tabular}{@{}lllll@{}}
\toprule
Parameters      & Ctl\_Frac\_KV & h48h\_Frac\_KV & h72h\_Frac\_KV & h96h\_Frac\_KV \\ \midrule
$E_2$               & 0.91    & 0.82    & 8.48    & 0.0038    \\
$\eta_{2}$            & 163.23   & 124.74    & 96.86    & 203.77 \\
$\nu$           & 0.95   & 0.80    & 0.99    & 0.99    \\
sse             & 1.4222E-05   & 9.1126E-05     & 9.7456E-06    & 5.1307E-05    \\
rsquare         & 0.9931
   & 0.9694    & 0.9973    & 0.97709    \\
dfe             & 56            & 56             & 46             & 55             \\
adjusted rsquare & 0.9942   & 0.9684    & 0.9972    & 0.9763    \\
rmse            & 0.0005   & 0.0013    & 0.0005    & 0.0010    \\ \bottomrule
\end{tabular}
\end{table}
\subsection{FC mechanical models quantify cell cycle changes during maturation}
In addition to quantifying viscoelastic changes due to pharmacological interventions, the fractional KV model also quantifies known cytoskeletal changes due to cell cycling and maturation of the cells, as shown in Fig. \ref{FigFC_Time}. 
Both the steady viscosity and the elastic modulus quantify changes as macrophages mature (Table \ref{TabFC_Time}). From the induction of differentiation to 72 hours after, there is a gradual reduction in steady state viscosity $\eta_{2}$. Interestingly, apoptosis (cell death) sets in around 96 hours following differentiation \cite{Ekpenyong2012e,Ekpenyong2012c}. The steady state viscosity jumps to a value greater than those of healthy cells. Succinctly, we have used fractional calculus models to further characterize inherent biophysical parameters of cells which are already known to be useful diagnostic parameters \cite{Guck2005}. In fact, since recent work also present cell and tissue mechanical properties as novel therapeutic targets in the case of cancer metastasis \cite{Prathivadhi-Bhayankaram2016d}, acute lung injury \cite{Ekpenyong2017b}, chronic pulmonary obstructive disease \cite{Ionescu2017}, the importance of using more robust and accurate models for extracting viscoelastic parameters of cells and tissues is easy to see. 
\section{Conclusion}
In this work, we have recapitulated some limitations of integer-order mechanical models in capturing features of compliance data from macrophages under various clinically relevant conditions. By comparing  fits using both integer order models and fractional calculus versions in Mittag-Leffler form, we found that the viscoelastic parameters from fractional Kelvin-Voigt model quantify the pharmacological interventions and maturation of macrophages more robustly than the integer-order models. Fractional calculus modeling therefore provides robust and more generally applicable biophysical characterization of cells in health and disease.

\bibliography{library}
\section*{Acknowledgments}
The authors are grateful to Prof Dr Jochen Guck of Max Planck Institute for the Science of Light, in whose Laboratory 
in the University of Cambridge, A. Ekpenyong, carried out all the compliance measurements modeled in this work.
Funding: This work was supported by Creighton University's College of Arts and Sciences startup grant to A. Ekpenyong (240133-215000 FY19) and by a Creighton University Ferlic Scholarship to A. Vo. 

\section*{Conflict of interest}
The authors declare there is no conflict of interest.
\end{document}